\documentclass[12pt]{article}%
\usepackage[slantedGreek]{mathpazo}
\usepackage{graphicx}
\usepackage{amsfonts}
\usepackage{setspace}
\usepackage{amssymb}
\usepackage{amsthm}
\usepackage{graphicx}
\usepackage{amsmath}%
\usepackage{color}
\usepackage{natbib}
\usepackage{float}
\usepackage[utf8]{inputenc}    
\usepackage[T1]{fontenc}       

\numberwithin{Def}{section}

\numberwithin{Cor}{section}

\title{Sequential Bayesian Learning for Merton's Jump Model with Stochastic Volatility}
\author{
	Eric Jacquier\thanks{Boston University, Questrom School of Business, email: jacquier@bu.edu}\\
	\texttt{Boston University, Questrom School of Business}
	\and
	Nicholas Polson\thanks{The University of Chicago Booth School of Business, email:nicholas.polson@chicagobooth.edu}\\
	\texttt{The University of Chicago Booth School of Business}
	\and
	Vadim Sokolov\thanks{George Mason University, Volgenau School of Engineering, email: vsokolov@gmu.edu}\\
	\texttt{George Mason University, Volgenau School of Engineering}
}

\date{First Draft: August 2014\\
	  This Draft: October 2016
	}
\graphicspath{{./fig/}}

\begin{document}

\maketitle
\begin{abstract}
\noindent 
Jump stochastic volatility models are central to
financial econometrics for volatility forecasting, portfolio risk management,
and derivatives pricing. Markov Chain Monte Carlo (MCMC) algorithms are computationally unfeasible for the sequential learning of volatility state variables and parameters,
whereby the investor must update all posterior and predictive densities as new information arrives. We develop a particle filtering and learning algorithm to sample posterior distribution in Merton's jump stochastic volatility. This allows to filter spot volatilities and jump times, together with sequentially updating (learning) of   jump and volatility parameters. We illustrate our methodology on Google's stock return. We conclude with directions for future research. 

\vspace{0.1in}
\end{abstract}
\newpage
\singlespacing

\section{Introduction}
Jump stochastic volatility models are central to many questions in finance such as pricing, or debt-and-credit risk assessment. \cite{merton1976option,duffie2000transform} provide theoretical treatments of derivatives pricing and \cite{merton1974pricing, korteweg2008volatility} provide applications to debt and credit risk assessment. Most theoretical treatments in the literature assume the availability of efficient estimates of volatility, jumps and parameters. Efficient estimates of the current volatility state and jump parameters are available from Markov Chain Monte Carlo (MCMC) algorithms, see \cite{johannes2010, jacquier2004bayesian}. A number of 
authors have analyzed jump diffusion models by MCMC, see \cite{eraker2003impact,li2008bayesian,fulop2012bayesian}. One caveat is that MCMC algorithms  are computationally demanding, and are not feasible  for sequential learning. Essentially, MCMC algorithm needs to be run every time new information is available.

Particle filtering (PF) and learning (PL) algorithms, on the other hand, efficiently incorporate new information into the parameter  learning process. Early PL algorithms  where plagued by degeneracy problem which hampered their performance. Particle learning (\cite{carvalho2010particle, warty2016}) algorithms deliver posterior
and predictive densities of parameters and latent variable as new information arrives. Sequential parameter learning is obtained by
tracking a state vector of conditional sufficient statistics. 

An important goal of an investor, for example, is to characterize the density of current and future
returns to draw inference on the riskiness of a portfolio, probability of shortfall, or value-at-risk. 
The distribution of future volatility is an input in the computation of derivative prices or their
hedge ratios. We provide a versatile model of access returns that combines Merton's pure jump formulation with stochastic
volatility. Within this model, the investor needs to learn about the state variables, namely, volatility, 
jump times and jump sizes, and the model parameters from the observed returns.  PL methods are particularly well-suited for empirical finance
applications for several reasons.

\begin{enumerate}
\item They are designed to be sequential, updating the relevant posterior distribution as new information 
(data) is obtained, with minimal computing resources.  Bayesian inference tools directly apply to these 
PF algorithms as they produce posterior or predictive densities relevant to the models used. 

\item Akin to MCMC algorithms, particle filtering and learning can be extended to simultaneously estimate both structural
parameters and latent variables. For example, one can separate out the effects of jumps and 
stochastic volatility in equity returns. 

\item As conditional simulation methods, they avoid optimization. From a practical perspective, 
PF and PL methods are therefore extremely fast in terms of computing time. This has many advantages, 
particularly for higher-dimensional multivariate models.
\end{enumerate}

One can also included option price information into the inference  problem, see for example \cite{polson2003,johannes2009particle,yun2014out}. Whilst we only account for stochastic jumps, it is 
easy to add deterministic jump components to account for example, for earnings announcement effects, see \cite{dubinsky2005}.

The rest of the paper is as follows. Section 2 provides a review of particle  filtering methods. 
Section 3 provides the main contribution of our paper and an algorithm for sequential filtering states 
and performing parameter learning for Merton Jump model together with stochastic volatility (\cite{jacquier1995models,jacquier2004bayesian}). Section 4 provide a application to Google's stock return. Finally, Section 5 
concludes with directions for future research.

\section{Particle Filtering for Merton's Jump Model}
Merton's Jump Stochastic volatility model  has a discrete time version for log-returns, $y_t$, with jump times, $J_t$, jump sizes, $Z_t$, and 
spot stochastic volatility, $V_t$, given by the dynamics
\begin{align*}
y_{t} & \equiv \log \left( S_{t}/S_{t-1}\right) =\mu + V_t \varepsilon_{t}+J_{t}Z_{t} \\
V_{t+1} & = \alpha_v + \beta_v V_t + \sigma_v \sqrt{V_t} \varepsilon_{t}^v
\end{align*}
where $ \mathbb{P} \left ( J_t =1 \right ) = \lambda $. The errors $(\varepsilon_{t},\varepsilon_{t}^v)$ 
are possibly correlated bivariate normals.

The investor must obtain optimal filters for $ (V_t,J_t,Z_t) $, and learn the posterior densities
of the parameters $ (\mu, \alpha_v, \beta_v, \sigma_v^2 , \lambda ) $. These estimates will be 
conditional on the information available at each time.

\subsection{Pure Jump Merton model}

Let $S_t$ denote a stock or asset price. We have  historical log-returns $ y^t = ( y_1 , \ldots , y_t )$ 
defined by $ y_t = \log ( S_t / S_{t-1} )$. Log-returns have a jump component as per
\begin{equation*}
y_{t} = \mu + \sigma \varepsilon_{t} + J_{t}Z_{t},
\end{equation*}%
with the probability of jump $\mathbb{P}\left[ J_{t}=1\right] = \lambda \in \left( 0,1\right) $ and
$Z_t$ is the jump size. The Merton model involves the following  hierarchical conditional 
distributions:
\begin{align*}
(\mu,\sigma^2) &	\sim 	NIG(m,n,a,b),\\
\lambda 		&\sim Beta(\alpha,\beta),\\
J_t|\Theta 		&\sim 	Ber(\lambda),\\
Z_t|\Theta,J_t &\sim 	\mathcal{N}(\mu_J,\sigma_J^2\sigma^2),\\
Y_t|\Theta,J_t,Z_t & \sim \mathcal{N}(\mu+J_tZ_t,\sigma^2),
\end{align*}
where $NIG$ denotes the standard conditionally conjugate Normal-Inverse-Gamma distribution. The parameters $\mu_J, \sigma_J^2$ are fixed to guarantee identification.
The goal of underlying dynamics is to provide sequential learning plot for states 
$(J_t,Z_t)$ and the unknown parameters $ (\mu, \sigma^2 , \mu_J,\sigma^2_J,\lambda )$. 

To construct a particle filtering and learning algorithm that samples from the set of Bayesian joint posterior distributions $p( J_t, Z_t , \theta | y^t )$, for $1\le t\le T$. We track a particle vector that tracks the 
hidden states $ \{ J_t , Z_t  \}^{(i)} $ and the conditional sufficient statistics $ \{ s_t \}^{(i)} $ 
for any static parameters that need to be learned. Then attach them in one vector  $ \{ J_t , Z_t , s_t \}^{(i)} $ for $ 1 \leq i \leq N $.

\vspace{1em}
Our Particle learning algorithm has four steps:

\begin{enumerate}
\item Determine the conditional sufficient statistics $ s_t $ for the parameters $ \theta = (\mu, \sigma^2 , \lambda ) $.
We  write $p(\theta | s_t) $ where $ s_t = ( m_t , n_t , a_t , b_t ,\alpha_t , \beta_t )$ with the following distributional assumptions
$$
p( \mu , \sigma^2 | J_{1:t} , Z_{1:t} , y_{1:t} ) \equiv
p( \mu , \sigma^2 | s_t ) \sim NIG \left ( m_t , n_t , a_t , b_t \right ) \; \; {\rm and} \; \; ( \lambda | s_t ) \sim Beta (\alpha_t , \beta_t ).
$$
This leads to a current posterior distribution, $ p( \mu,\sigma^2, \lambda | s_t)$ that is proportional to 
$$(\sigma^2)^{-a_t-3/2} \exp \left( -\frac{n_t(\mu-m_t)^2 + 2b_t}{2\sigma^2} \right)
\lambda^{\alpha_t-1} (1-\lambda)^{\beta_t-1}. $$
The conditional likelihood $ p( Y_{t+1} , J_{t+1}, Z_{t+1} | \mu, \sigma^2 , \lambda ) $ 
 proportional to 
$$\lambda^{J_{t+1}} (1-\lambda)^{1- J_{t+1}} \times (\sigma_J^2 \sigma^2)^{-1/2} \exp \left( -\frac{(Z_{t+1}-\mu_J)^2 }{2\sigma_J^2 \sigma^2} \right) \times$$
$$(\sigma^2)^{-1/2} \exp \left( -\frac{(Y_{t+1}-\mu-J_{t+1}Z_{t+1})^2 }{2\sigma^2} \right).$$
Combining these two terms, leads to an updated conditional posterior, $  p( \mu,\sigma^2, \lambda | s_{t+1})$ with is again a $NIG \times Beta$ 
distribution with updated hyperparameters 
$$m_{t+1} = \frac{n_t m_t + R_{t+1}}{n_t+1}, n_{t+1} = n_t + 1, $$
$$ a_{t+1} = a_t + 1 \; {\rm and} \; 
b_{t+1} = b_t + \frac{n_t (R_{t+1}-m_t)^2}{2(n_t+1)} + \frac{(Z_{t+1}-\mu_J)^2}{2\sigma_J^2}$$ 
where $R_t = Y_t - J_tZ_t$, and $\alpha_{t+1} = \alpha_t + J_{t+1}$ and $\beta_{t+1} = \beta_t + 1 - J_{t+1}$.

\item Calculate the marginal predictive distribution, denoted by $p(y_{t+1}\mid J_t,Z_t,s_t)$ of the next asset return $y_{t+1} $ given the states $ J_{t+1} , Z_{t+1} $.
This a mixture distribution of the form
$$
\int \sum_{J_{t+1}} p( y_{t+1} | J_{t+1} , Z_{t+1}, \mu , \sigma^2 ) p( Z_{t+1} | \lambda )
p( J_{t+1} | \lambda ) p( \mu , \sigma^2 , \lambda | s_t ) d \mu d \sigma^2 d \lambda
$$
We marginalise out the states $ Z_{t+1} , J_{t+1} $ and the parameters $\mu, \sigma^2,\lambda$, as follows. First, marginalizing $Z_{t+1}$ and $\lambda$, we obtain
$$p(\mu,\sigma,J_{t+1},y_{t+1}|s_t) \propto \gamma_{t+1}^{1/2} B(\alpha_{t+1},\beta_{t+1}) \times NIG(m_{t+1},n_{t+1},a_{t+1},b_{t+1})$$ where with hyperparameter updates are given by
$$\gamma_{t+1} = \frac{1}{1+J_{t+1}\sigma_J^2},$$
$$m_{t+1} = \frac{n_tm_t+\gamma_{t+1}Q_{t+1}}{n_t + \gamma_{t+1}} \; {\rm where} \; Q_{t+1} = y_{t+1}-J_{t+1}\mu_J,$$
$$n_{t+1} = n_t + \gamma_{t+1},$$
$$a_{t+1} = a_t + \frac{1}{2},$$
$$b_{t+1} = b_t + \frac{n_t\gamma_{t+1}(Q_{t+1}-m_t)^2}{2(n_t+\gamma_{t+1})}.$$
Therefore, we have a marginal joint posterior 
$$
p( y_{t+1} , J_{t+1} | s_t ) \propto B(\alpha_{t+1},\beta_{t+1}) \gamma_{t+1}^{1/2} n_{t+1}^{-1/2} b_{t+1}^{-a_{t+1}}.
$$
Finally, marginalizing over the jump times $J_{t+1}$ leads to the require predictive for resampling
$$
p( y_{t+1} | s_t ) \propto \sum_{J_{t+1}} p( J_{t+1}, y_{t+1} | s_t )
$$
where $ s_t$ is the current conditional sufficient statistic.
Given the next observation $y_{t+1}$, we then use it to resample particles $ \{ J_t , Z_t , s_t \}^{(i)} $ 

\item Propagate a new state $J_{t+1} , Z_{t+1} $ using
the conditional posterior
$$
p( J_{t+1} , Z_{t+1} | s_t^{k(i)} , y_{t+1} ) \; .
$$
This is a mixture distribution where we can 
use an intermediate parameter vector draw from $ p( \theta | s_{t+1} ) $. In this manner, we obtian a new draw  $ ( J_{t+1} , Z_{t+1} )^{(i)} $
of the filtered distribution on states given the current return history.

To summarize, we can do this as follows:
\begin{enumerate}
\item Generate $p(J_{t+1}|s_t,y_{t+1})$.
\item Generate $p(\mu,\sigma^2|J_{t+1},s_t,y_{t+1})$.
\item Sample $Z_{t+1}|\mu,\sigma^2,J_{t+1},s_t,y_{t+1} \sim \mathcal{N}(\gamma_{t+1}\mu_J + (1-\gamma_{t+1})(y_{t+1}-\mu), \gamma_{t+1}\sigma_J^2\sigma^2)$.
\end{enumerate}

\item Finally, use the  update rule for sufficient statistics $ s_{t+1}^{(i)} = \mathbb{S} \left ( s_t^{k(i)} , (J_{t+1} , Z_{t+1})^{(i)} \right )$.

Here we have used the \emph{resampled} $s_t^{k(i)}$ in the update and the sampled $ ( J_{t+1} , Z_{t+1} )^{(i)} $ from the previous step.
\end{enumerate}

\subsection{Extension to Jumps with Stochastic Volatility}

We now  assume that the log-returns not only have a jump component but also a 
stochastic volatility component, with dynamics given by
\[
\begin{aligned}
Y_t & = \log(S_t/S_{t-1}) = \mu + \sqrt{V_t}\epsilon_t + J_tZ_t\\
V_{t+1} &  = \alpha_v + \beta_vV_t + \sigma_v\sqrt{V_t}\sigma_t^V
\end{aligned}
\]
The relevant  conditional distributions are now given by hierarchical structure of conditional distributions, given by 

\begin{align*}
Z_t|\Theta,J_t, V_t & \sim \mathcal{N}(\mu_J,\sigma_J^2),\\
Y_t|\Theta,J_t,Z_t & \sim \mathcal{N}(\mu+J_tZ_t,\sqrt{V_t}\epsilon_t)\\
(\mu_J,\sigma_J^2) & \sim fixed\\
(\alpha_v,\beta_v) & \sim fixed\\
\mu 	& \sim \mathcal{N}(m,n^{-1}),\\
\lambda 		& \sim Beta(\alpha,\beta),\\
J_t|\Theta 	& \sim Ber(\lambda),\\
(\alpha_v, \beta_v, \sigma_v^2) & \sim NIG(m_v, n_v, a_v, b_v)\\
V_{t+1} 	& \sim \mathcal{TN} \left ( (1-\beta_v)\alpha_v + \beta_v V_t , \sigma_v^2 V_t \right )
\end{align*}
where $ \mathcal{TN}$ denotes a truncated normal distribution.

\begin{figure}[H]
\centering
\includegraphics[width=0.5\linewidth]{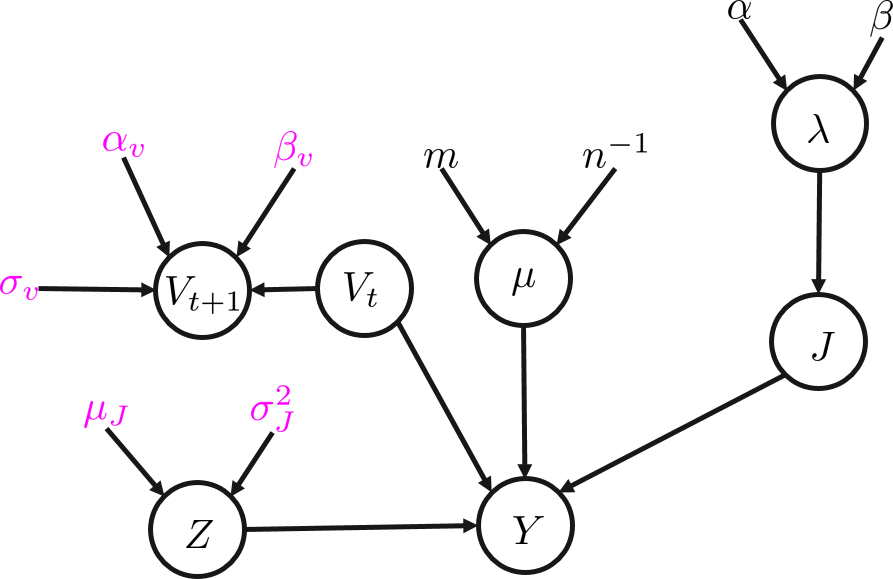}
\caption{Jusm Stochastic Volatility Model as a Graphical model}
\label{fig:graph-model}
\end{figure}

The hierarchical conditional independences structure can also be represented as a graphical model as shown in Figure~\ref{fig:graph-model}
We see, that the jump sizes depend on the magnitude of the current volatility $V_t$. Thus, the state variables $(J_t, Z_t, V_t)$ are defined by two parameters $\Theta = (\mu, \lambda)$. At the same time the distribution of $\Theta$ is defined by the hyperparameters that have sufficient statistics $s_t = (m_t,n_t,\alpha_t, \beta_t)$, which we denote by $\Theta \sim p(\Theta | s_t)$. We now construct the conditional posterior distribution of the parameter $\mu$. Under a sufficient statistic structure, we have  
$$
p( \mu | J_{1:t} , Z_{1:t} , y_{1:t} ) \equiv
p( \mu  | s_t ) \sim \mathcal{N} \left ( m_t , n_t \right ) \; \; {\rm and} \; \; ( \lambda | s_t ) \sim Beta (\alpha_t , \beta_t )
$$
Thus each particle $i$ is defined by $\left\{J_t, Z_t, V_t, s_t\right\}^{(i)}$, for $1 \le i \le N$. Let $\zeta$ denote joint variable $(J_{t+1},V_{t+1}, s_t)$.

The first step is to find the posterior distributions for $\mu$ and $\lambda$. Given $\lambda \sim Beta(\alpha, \beta)$, the posterior is given by
\[
\lambda | J_{t+1}, s_t \sim Beta(\alpha_t + J_{t+1}, \beta_t +1 - J_{t+1})
\]
with an update value for hyper parameters
\begin{equation*}\label{eqn:lupdate}
\alpha_{t+1} = \alpha_t + J_{t+1}, \ \beta_{t+1} = \beta_t + 1 - J_{t+1}
\end{equation*}
Now we look at $p(Y_t|Z_t,J_t,\mu, \lambda)$. Note, that $J_tZ_t \sim N(J_t\mu_J, J_t\sigma_J)$ and independent of $\epsilon_t$. Thus we can easily marginalize $Z_t$ using sum of normal variables formula, and obtain
\begin{align*}
y_{t} = &\mu + J_t \mu_J + \sqrt{J_t\sigma_J^2 + V_{t}}\epsilon_t\\
Y_t | \Theta,J_t \sim & N(\mu + J_t \mu_J, \sqrt{J_t\sigma_J^2 + V_{t}})
\end{align*}

We denote the precision parameter by $\gamma_t = \dfrac{1}{J_t\sigma_J^2 + V_{t}}$ and $Q_{t+1} = y_{t+1} - J_{t+1}\mu_J$.

Now, the joint distribution for $\mu$ and $y_{t+1}$ is given by 
\[
p(y_{t+1}, \mu|\zeta_t) = p(y_{t+1} | \mu,\zeta_t)p(\mu| s_t) = p(\mu|y_{t+1},\zeta_t)p(y_{t+1}|\zeta_t)
\]

\[
 p(y_{t+1} | \mu,\zeta_t) = \frac{\sqrt{\gamma _{t+1}} \exp \left(-\frac{1}{2} \gamma _{t+1} \left(\mu - Q_{t+1}\right){}^2\right)}{\sqrt{2 \pi }}
\]
and 
\[
p(\mu|s_t) = \frac{\sqrt{n_t} \exp\left(-\frac{1}{2} n_t \left(\mu -m_t\right){}^2\right)}{\sqrt{2 \pi }}
\]

\[
p(y_{t+1}, \mu|\zeta_t) \propto  \exp \left(-\frac{1}{2} \gamma _{t+1} \left(\mu - Q_{t+1}\right){}^2\right)\exp\left(-\frac{1}{2} n_t \left(\mu -m_t\right){}^2\right).
\]
On the other hand
\[
 p(\mu|y_{t+1},\zeta_t)p(y_{t+1}|\zeta_t) \propto \exp\left(-\frac{1}{2} n_{t+1} \left(\mu -m_{t+1}\right){}^2\right)\exp\left(-\frac{1}{2} \tau_{t+1}\left(\mu - Q_{t+1}\right){}^2\right)
\]
To calculate mean and standard deviation of  the predictive likelihood $p(y_{t+1}|\zeta_t)$ and posterior for mean $p(\mu|y_{t+1},\zeta_t)$, we use the identity

\begin{align*}
&\frac{\left(x-\mu _1\right){}^2}{\sigma _1}+\frac{\left(x-\mu _2\right){}^2}{\sigma _2} = \frac{\left(\mu _1-\mu _2\right){}^2}{\sigma _1+\sigma _2}+\frac{\left(x-\mu _3\right){}^2}{\sigma _3}\\
&\mu _3\ =  \sigma _3 \left(\frac{\mu _1}{\sigma _1}+\frac{\mu _2}{\sigma _2}\right),~\sigma _3 =  \frac{1}{\frac{1}{\sigma _2}+\frac{1}{\sigma _1}}.
\end{align*}

We apply this identity with correspondence $\mu_1 \rightarrow Q_{t+1} = y_{t+1} - J_{t+1}\mu_J$, $\mu_2 \rightarrow m_t$, $\mu_3 \rightarrow m_{t+1}$, $\sigma_1 \rightarrow \gamma_{t+1}^{-1}$, $\sigma_2 \rightarrow n_t^{-1}$, $\sigma_3 \rightarrow n_{t+1}^{-1}$, and $\sigma_1 + \sigma_2 \rightarrow \tau_{t+1}^{-1}$.

From this we can calculate the updates for the hyper parameters as follows
\begin{equation*}\label{eqn:muupdate}
m_{t+1} = \frac{n_t}{n_t + \gamma_{t+1}}m_t +  \frac{\gamma_{t+1}}{n_t + \gamma_{t+1}}Q_{t+1} = \frac{n_tm_t + \gamma_{t+1}Q_{t+1}}{n_t + \gamma_{t+1}}, \ n_{t+1} = n_t + \gamma_{t+1},
\end{equation*}
The predictive distribution is given by
\[
p(y_{t+1}|\zeta_t) \propto \exp \left(-\frac{\left(Q_{t+1} - m_t\right){}^2}{2 \left(\frac{1}{n_t}+\frac{1}{\gamma _{t+1}}\right)}\right).
\]
The likelihood precision is $\tau_{t+1} = \frac{n_t \gamma_{t+1}}{n_t + \gamma_{t+1}}$,  and by normalizing, we obtain
\[
p(y_{t+1}|\zeta_t) = \phi(J_{t+1}\mu_J + m_t,\tau_{t+1}^{-1}).
\]
The predictive likelihood $p(y_{t+1} |V_{t+1}, s_t)$ is determined by summing out $J_{t+1}$ as
\begin{equation}\label{eqn:pred}
p(y_{t+1} |V_{t+1}, s_t) = \sum_{J_{t+1} \in \{0,1\}}p(y_{t+1} | J_{t+1}, V_{t+1}, s_t)p(J_{t+1}|s_t),
\end{equation}
where
\begin{align*}
 p(J_{t+1} = 0|s_t)& = \frac{\alpha_t}{\alpha_t+\beta_t} \mbox{ and }
 p(J_{t+1} = 1|s_t) =  \frac{\beta_t}{\alpha_t + \beta_t}\\ 
 p(y_{t+1} | J_{t+1}=0, V_{t+1}, s_t) & =  \phi( m_t,n_t^{-1} + V_{t+1}) \\
 p(y_{t+1} | J_{t+1}=1, V_{t+1}, s_t) & =\phi(\mu_J + m_t,n_t^{-1} + \sigma_J^2+ V_{t+1}).
\end{align*}

Finally, we need to find formulas for propagating the state vector $(J_t, Z_t, V_t)$, given the latest observation $y_{t+1}$. Note, that $V_{t+1}$ is conditionally independent of $y_{t+1}$, given $V_t$. Thus, we can propagate the volatility variable by draying from the truncated normal $ V_{t+1} \sim \mathcal{TN} \left ( \alpha_v + \beta_v V_t , \sigma_v \sqrt{V_t} \right )$.

The odds ratio for $J_{t+1}$, is given by
\begin{equation*}\label{eqn:jprop}
\begin{aligned}
\frac{p(J_{t+1} = 0 | y_{t+1}, V_{t+1}, s_t)}{p(J_{t+1} = 1 | y_{t+1}, V_{t+1}, s_t)} &= \frac{p(y_{t+1} | J_{t+1} = 0, V_{t+1}, s_t)}{p(y_{t+1} | J_{t+1} = 1, V_{t+1}, s_t)}\frac{p(J_{t+1} = 0 | s_t)}{p(J_{t+1} = 1 | s_t)}\\ &= 
\frac{ \phi( m_t,n_t^{-1} + V_{t+1})}{\phi(\mu_J + m_t,n_t^{-1} + \sigma_J^2+ V_{t+1})}\frac{\alpha_t}{\beta_t}
\end{aligned}
\end{equation*}

If we sample $J_{t+1} = 0$, then size of the jump $Z_{t+1}$ is irrelevant, otherwise we have
\[
Z_{t+1} = y_{t+1} - m_t + \sqrt{n^{-1}_t + \sigma_J^2 + V_{t+1}}\epsilon_t
\]
Thus, the posteriors is
\begin{equation*}\label{eqn:zprop}
p(Z_{t+1} | y_{t+1}, J_{t+1} = 1, V_{t+1}, s_t)  = \phi(y_{t+1} - m_t, n^{-1}_t + \sigma_J^2 + V_{t+1})
\end{equation*}

This leads to the following algorithm
\begin{enumerate}
\item Resample index $k_{t+1}^{(i)} \sim Mult\left\{p\left(y_{t+1}| (V_{t+1},s_t)^{(i)}\right)\right\}$, so that $(V_{t+1},s_t)^{k_{t+1}^{(i)}} \sim p(V_{t}, s_t | y_{1:t+1})$

\item Propagate $Z_t^{k_{t+1}^{(i)}}$ and $J_t^{k_{t+1}^{(i)}}$, using equations (\ref{eqn:jprop}) and (\ref{eqn:zprop}) correspondingly

\item Update sufficient statistics $s_{t+1}^{i} = S\left(s_t^{k_{t+1}^{(i)}}\right)$, using equations (\ref{eqn:lupdate}) and $\ref{eqn:muupdate}$. 

\item Propagate  volatilities by drawing $ V_{t+2} \sim \mathcal{TN} \left ( (1- \beta_v)\alpha_v + \beta_v V_{t+1} , \sigma_v \sqrt{V_{t+1}} \right )$

\end{enumerate}

\section{Application: Google stock returns}
We now implement our particle filtering and learning algorithm. Namely, we applied the algorithm to the 1929 Google and S\&P500 daily stock returns from January 3 2007 to August 29, 2014. 
Table \ref{table:return-summary} provides summary statistics for the returns. 
\begin{table}[H]
\centering
\begin{tabular}{ccccccc}
	\hline
	        &   Mean   & Std. Deviation & Skewness & Kurtosis &   Min   &  Max  \\ \hline
	Google  & 4.84e-04 &     0.0193     &   0.21   &   8.7    & -0.123  & 0.161 \\ \hline
	S\&P500 & 1.78e-04 &     0.0135     &  -0.383  &   9.35   & -0.0911 & 0.101 \\ \hline
\end{tabular} 
\caption{Summary statistics for the daily log-returns}
\label{table:return-summary}
\end{table}
As expected, the daily returns exhibit a large amount of kurtosis consistent with time varying
second moments, as in stochastic volatility or jumps.  Figure \ref{fig:returns-prices} below shows the price and return data for the selected period, confirming the time variation in volatility.
\begin{figure}[H]
\begin{tabular}{cc}
\includegraphics[width=0.5\linewidth]{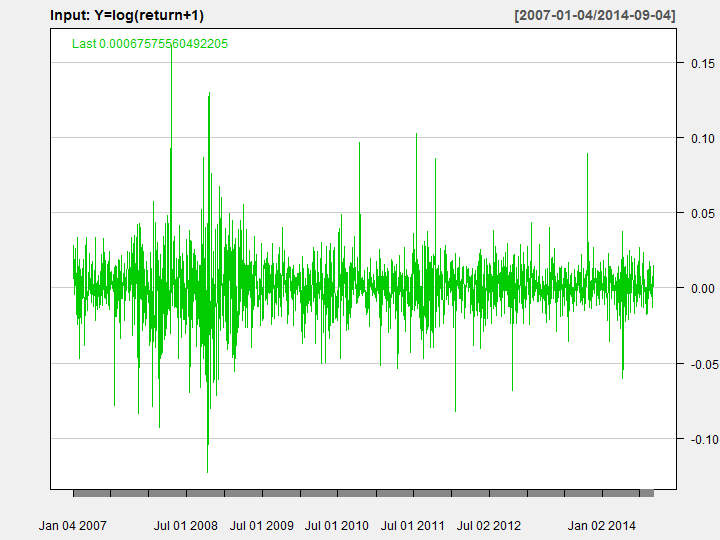} & \includegraphics[width=0.5\linewidth]{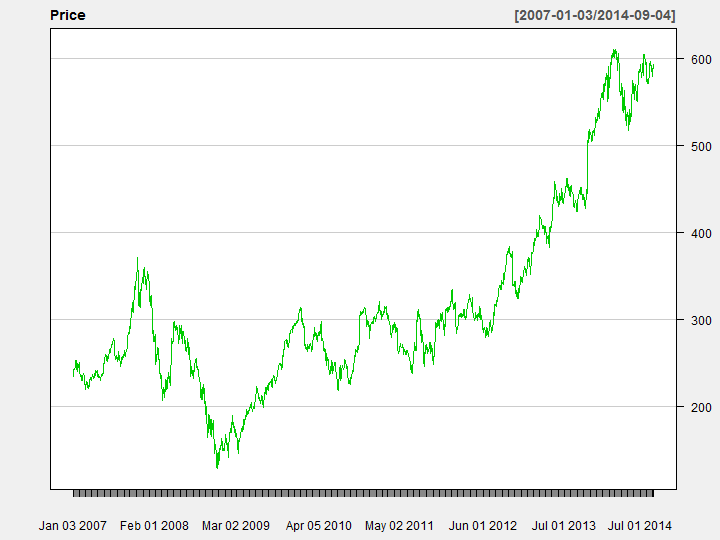}\\
(a) GOOGLE Daily Log-returns   & (b)  GOOGLE Daily Price\\
\includegraphics[width=0.5\linewidth]{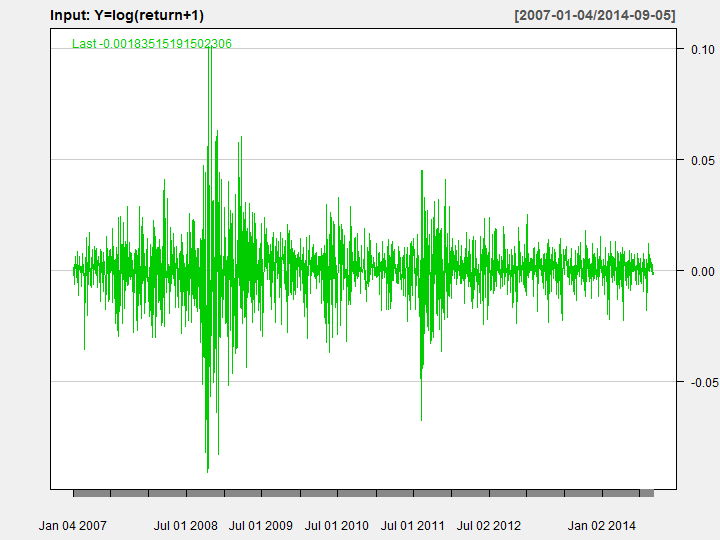} & \includegraphics[width=0.5\linewidth]{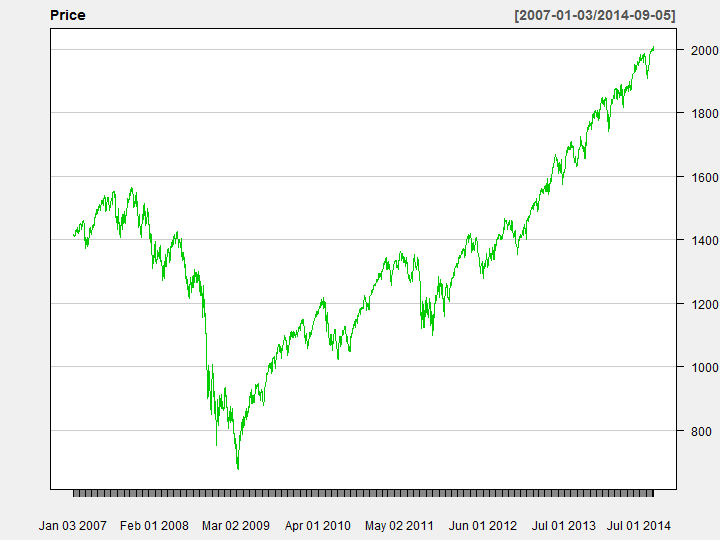}\\
(c) S\&P500 Daily Log-returns   & (d)  S\&P500 Daily Price
\end{tabular}
\caption{Price and return data for Google stock}\label{fig:returns-prices}
\end{figure}

Tables \ref{table:parameters} and \ref{table:parameters-sp} provides posterior means and standard deviation of the parameters and state variables for the models estimated, as well as prior values. 
The second and fourth columns provide parameter posterior means, and standard deviations in parentheses, for the Merton and Merton-SV models.
\begin{table}[H]
\begin{tabular}{lllll}
	             & Prior (Merton) & Posterior (Merton)  & Prior (Merton-SV) & Posterior (Merton-SV) \\ \hline
	$\mu_J$      & -0.04          &                     & -0.04             &  \\
	$\sigma^2_J$ & 1              &                     &                   &  \\
	$\mu$        & 0.003          & 3.86e-03 (9.52e-04) & 0.003             & 1.57e-03 (2.96e-03)   \\
	$\sigma^2$   & 0.001          & 1.49e-04 (3.97e-05) & 0.01              & 1.35e-04 (7.94e-04)   \\
	$\lambda$    & 0.09           & 0.086(0.219)        & 0.5               & 7.15e-03 (0.0266)     \\
	$Z$          &                & -0.0399 (8.62e-03)  &                   & -0.0397 (0.0413)      \\
	$\alpha_v$   &                &                     & 0.0016            &  \\
	$\beta_v$    &                &                     & 0.99              &  \\
	$V$          &                &                     & 0.3               & 0.237 (0.0336)        \\ \hline
\end{tabular} 
\caption{Parameter Estimates and Priors fro Google}\label{table:parameters}
\end{table}
\begin{table}[H]
\begin{tabular}{lllll}
	             & Prior (Merton) & Posterior (Merton)  & Prior (Merton-SV) & Posterior (Merton-SV) \\ \hline
	$\mu_J$      & -0.04          &                     & -0.04             &  \\
	$\sigma^2_J$ & 1              &                     &                   &  \\
	$\mu$        & 0.003          & 2.77e-03(7.8e-04)   & 0.003             & 3.62e-04 (1.98e-03)   \\
	$\sigma^2$   & 0.001          & 7.05e-05 (2.62e-05) & 0.01              & 1.3e-04 (7.4e-04)     \\
	$\lambda$    & 0.09           & 0.0627 (0.213)      & 0.5               & 0.011 (0.0542)        \\
	$Z$          &                & -0.0398 (1.93e-03)  &                   & -0.0417 (0.0395)      \\
	$\alpha_v$   &                &                     & 0.0016            &  \\
	$\beta_v$    &                &                     & 0.99              &  \\
	$V$          &                &                     & 0.3               & 0.216 (0.0374)        \\ \hline
\end{tabular} 
\caption{Parameter Estimates and Priors fro S\&P 500}\label{table:parameters-sp}
\end{table}

\begin{figure}[H]
\centering
\begin{tabular}{cc}
 \includegraphics[width=0.5\linewidth]{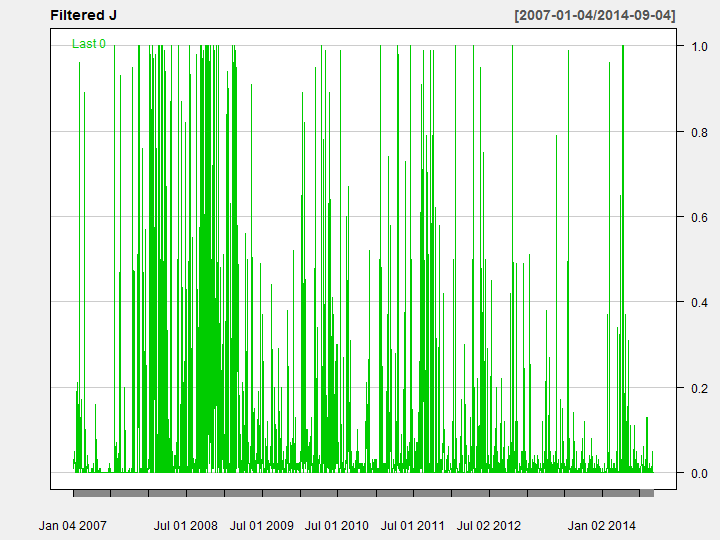}& \includegraphics[width=0.5\linewidth]{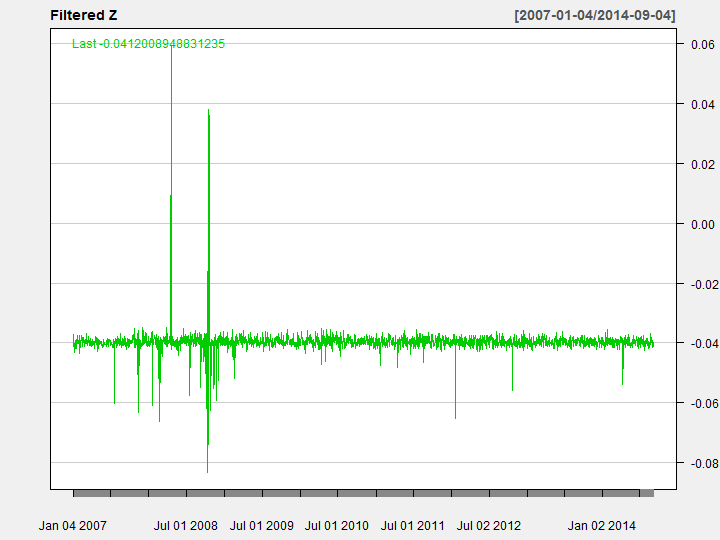} \\ 
 (a) Filtered J for GOOGLE & (b) Filtered Z for GOOGLE\\
  \includegraphics[width=0.5\linewidth]{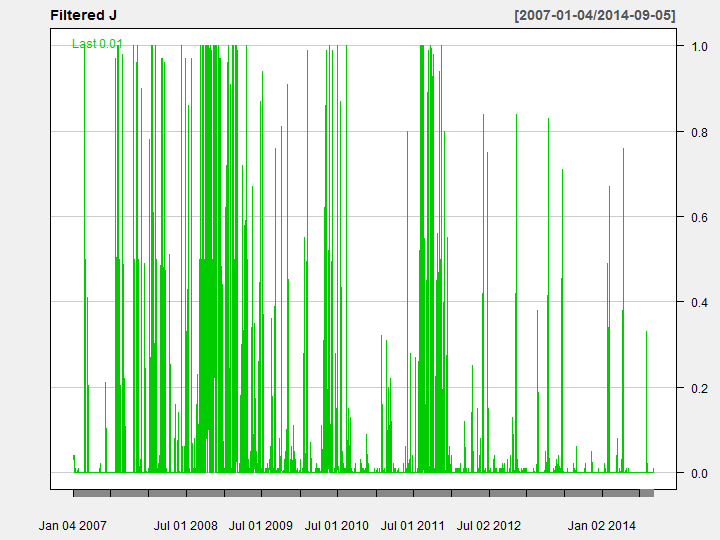}& \includegraphics[width=0.5\linewidth]{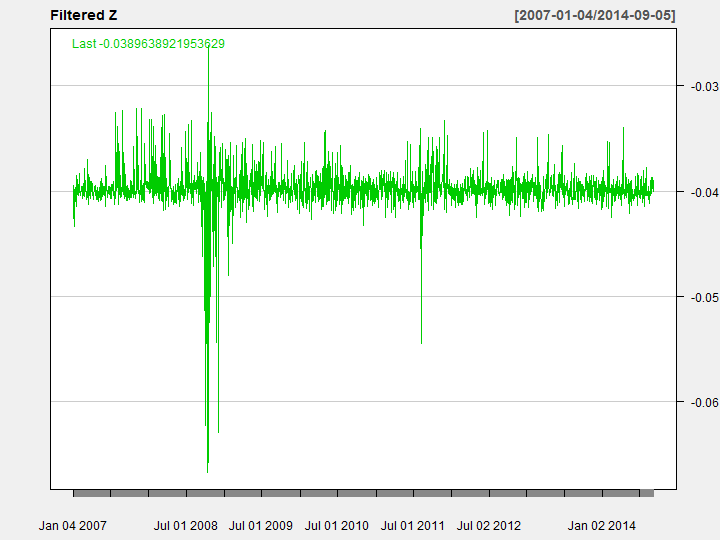} \\ 
  (a) Filtered J for S\&P500 & (b) Filtered Z for S\&P500
\end{tabular} 
\caption{Value of the filtered state variables for Merton model}
\label{fig:filered-jz-merton}
\end{figure}

Consider the pure Merton model without stochastic volatility. Figure \ref{fig:filered-jz-merton} shows the filtered state parameters $Z$ ans $J$ estimated using the Merton jump model. We can see that the cluster of jumps in July 2008. Such a clustering indicates that the model may be misspecified. One reason for that is that the volatility is fixed in the model and thus, all of the large moves on returns are attributed to jumps. However, the clustering of jumps is extremely unlikely in reality due to the i.i.d assumption on the jump time and size specification and infrequent nature of jumps. 

These clustered jumps of 2008 in fact reflect higher volatility of returns.


The fourth column of Tables \ref{table:parameters} and \ref{table:parameters-sp} provides the parameters estimates for the Merton-SV model. Adding stochastic volatility to the jump model has the expected effect of reducing the amount of jumps in the returns. Virtually, all of the changes on returns are explained by the stochastic volatility 



\section{Discussion}

Particle filtering methods are flexible and fast to compute. They provide a simple solution to the sequential inference problem where Markov Chain Monte Carlo (MCMC) are computationally expensive as they have to be re-run every time a new data point arrives.
We develop and implement a particle filtering and learning algorithm that provides full inference for Merton's jump stochastic volatility model.
To perform sequential parameter learning we exploit a conditional sufficient statistic state variable that we filter with particle methods and then we draw parameters in an off line fashion. This provides an efficient approach to parameter inference.

There are a number of possible extensions of our work. On the finance side, there are many models with a similar nature to Merton's original specification such as the \cite{leland1996optimal} model of corporate credit. These models are state space-models and are amenable to particle filtering methods. On the econometrics side, extensions to continuous-time jump diffusions (\cite{johannes2009optimal} with
infinite activity jumps (\cite{li2008bayesian}) or to self-exciting jump processes (\cite{fulop2012bayesian,ait2009estimating,ait2012analyzing}) is an avenue for future study.  Incorporating a leverage effect (or correlated errors) together with multivariate models is another area of interest (\cite{jacquier1995models,jacquier2004bayesian}). We leave these as avenues for future study. 

We intent to extend our analysis to currencies (FX) and security portfolios (e.g. market indexes). Furthermore, we will demonstrate both computational accuracy of estimation  differences between particle filters and MCMC based methods.

\bibliography{ref}

\begin{thebibliography}{}

\bibitem[\protect\astroncite{A{\"\i}t-Sahalia and
  Jacod}{2009}]{ait2009estimating}
A{\"\i}t-Sahalia, Y. and J.~Jacod\leavevmode\nopagebreak\newline 2009.
\newblock Estimating the degree of activity of jumps in high frequency data.
\newblock {\em The Annals of Statistics}, Pp.~ 2202--2244.

\bibitem[\protect\astroncite{A{\"\i}t-Sahalia and
  Jacod}{2012}]{ait2012analyzing}
A{\"\i}t-Sahalia, Y. and J.~Jacod\leavevmode\nopagebreak\newline 2012.
\newblock Analyzing the spectrum of asset returns: {{Jump}} and volatility
  components in high frequency data.
\newblock {\em Journal of Economic Literature}, 50(4):1007--1050.

\bibitem[\protect\astroncite{Carlin et~al.}{1992}]{carlin1992monte}
Carlin, B.~P., N.~G. Polson, and D.~S. Stoffer\leavevmode\nopagebreak\newline
  1992.
\newblock A {{Monte Carlo}} approach to nonnormal and nonlinear state-space
  modeling.
\newblock {\em Journal of the American Statistical Association},
  87(418):493--500.

\bibitem[\protect\astroncite{Carpenter et~al.}{1999}]{carpenter1999improved}
Carpenter, J., P.~Clifford, and P.~Fearnhead\leavevmode\nopagebreak\newline
  1999.
\newblock Improved particle filter for nonlinear problems.
\newblock {\em IEE Proceedings-Radar, Sonar and Navigation}, 146(1):2--7.

\bibitem[\protect\astroncite{Carvalho et~al.}{2010}]{carvalho2010particle}
Carvalho, C., M.~S. Johannes, H.~F. Lopes, and
  N.~Polson\leavevmode\nopagebreak\newline 2010.
\newblock Particle learning and smoothing.
\newblock {\em Statistical Science}, 25(1):88--106.

\bibitem[\protect\astroncite{Dubinsky and Johannes}{2005}]{dubinsky2005}
Dubinsky, A. and M.~Johannes\leavevmode\nopagebreak\newline 2005.
\newblock Earnings announcements and equity options.
\newblock {\em Working Paper}.

\bibitem[\protect\astroncite{Duffie}{1996}]{duffie1996state}
Duffie, D.\leavevmode\nopagebreak\newline 1996.
\newblock State-space models of the term structure of interest rates.
\newblock In {\em Stochastic {{Analysis}} and {{Related Topics V}}}, Pp.~
  41--67.
\newblock {Springer}.

\bibitem[\protect\astroncite{Duffie et~al.}{2000}]{duffie2000transform}
Duffie, D., J.~Pan, and K.~Singleton\leavevmode\nopagebreak\newline 2000.
\newblock Transform analysis and asset pricing for affine jump-diffusions.
\newblock {\em Econometrica}, 68(6):1343--1376.

\bibitem[\protect\astroncite{Eraker et~al.}{2003}]{eraker2003impact}
Eraker, B., M.~Johannes, and N.~Polson\leavevmode\nopagebreak\newline 2003.
\newblock The impact of jumps in volatility and returns.
\newblock {\em The Journal of Finance}, 58(3):1269--1300.

\bibitem[\protect\astroncite{Fulop et~al.}{2012}]{fulop2012bayesian}
Fulop, A., J.~Li, and J.~Yu\leavevmode\nopagebreak\newline 2012.
\newblock Bayesian learning of impacts of self-exciting jumps in returns and
  volatility.
\newblock {\em Working Paper}.

\bibitem[\protect\astroncite{Gordon et~al.}{1993}]{gordon1993novel}
Gordon, N.~J., D.~J. Salmond, and A.~F. Smith\leavevmode\nopagebreak\newline
  1993.
\newblock Novel approach to nonlinear/non-{{Gaussian Bayesian}} state
  estimation.
\newblock In {\em {{IEE Proceedings F}}-{{Radar}} and {{Signal Processing}}},
  volume 140, Pp.~ 107--113. {IET}.

\bibitem[\protect\astroncite{Jacquier et~al.}{2004}]{jacquier2004bayesian}
Jacquier, E., N.~Polson, and P.~Rossi\leavevmode\nopagebreak\newline 2004.
\newblock Bayesian {{Inference}} for {{SV}} models with {{Correlated Errors}}.
\newblock {\em Journal of Econometrics}.

\bibitem[\protect\astroncite{Jacquier et~al.}{1995}]{jacquier1995models}
Jacquier, E., N.~G. Polson, and P.~E. Rossi\leavevmode\nopagebreak\newline
  1995.
\newblock Models and priors for multivariate stochastic volatility.
\newblock Working paper, University of Chicago.

\bibitem[\protect\astroncite{Johannes and Polson}{2009}]{johannes2009particle}
Johannes, M. and N.~Polson\leavevmode\nopagebreak\newline 2009.
\newblock Particle filtering.
\newblock In {\em Handbook of {{Financial Time Series}}}, Pp.~ 1015--1029.
\newblock {Springer}.

\bibitem[\protect\astroncite{Johannes and Polson}{2010}]{johannes2010}
Johannes, M. and N.~Polson\leavevmode\nopagebreak\newline 2010.
\newblock {{MCMC Methods}} for {{Continuous}}-{{Time Financial Econometrics}}.
\newblock In {\em Handbook of {{Financial Econometrics}}: {{Applications}}},
  L.~P. HANSEN and Y.~A{\"I}T-SAHALIA, eds., volume~2 of {\em Handbooks in
  Finance}, Pp.~ 1--72.
\newblock San Diego: {Elsevier}.

\bibitem[\protect\astroncite{Johannes et~al.}{2009}]{johannes2009optimal}
Johannes, M.~S., N.~G. Polson, and J.~R. Stroud\leavevmode\nopagebreak\newline
  2009.
\newblock Optimal filtering of jump diffusions: {{Extracting}} latent states
  from asset prices.
\newblock {\em Review of Financial Studies}, 22(7):2759--2799.

\bibitem[\protect\astroncite{Korteweg and
  Polson}{2008}]{korteweg2008volatility}
Korteweg, A. and N.~Polson\leavevmode\nopagebreak\newline 2008.
\newblock Volatility, {{Liquidity}}, {{Credit Spreads}} and {{Bankruptcy
  Prediction}}.
\newblock Technical report, Stanford University, Working Paper.

\bibitem[\protect\astroncite{Leland and Toft}{1996}]{leland1996optimal}
Leland, H.~E. and K.~B. Toft\leavevmode\nopagebreak\newline 1996.
\newblock Optimal capital structure, endogenous bankruptcy, and the term
  structure of credit spreads.
\newblock {\em The Journal of Finance}, 51(3):987--1019.

\bibitem[\protect\astroncite{Li et~al.}{2008}]{li2008bayesian}
Li, H., M.~T. Wells, and L.~Y. Cindy\leavevmode\nopagebreak\newline 2008.
\newblock A {{Bayesian}} analysis of return dynamics with {{L{\'e}vy}} jumps.
\newblock {\em Review of Financial Studies}, 21(5):2345--2378.

\bibitem[\protect\astroncite{Merton}{1974}]{merton1974pricing}
Merton, R.~C.\leavevmode\nopagebreak\newline 1974.
\newblock On the pricing of corporate debt: {{The}} risk structure of interest
  rates.
\newblock {\em The Journal of finance}, 29(2):449--470.

\bibitem[\protect\astroncite{Merton}{1976}]{merton1976option}
Merton, R.~C.\leavevmode\nopagebreak\newline 1976.
\newblock Option pricing when underlying stock returns are discontinuous.
\newblock {\em Journal of financial economics}, 3(1-2):125--144.

\bibitem[\protect\astroncite{Pitt and Shephard}{1999}]{pitt1999filtering}
Pitt, M.~K. and N.~Shephard\leavevmode\nopagebreak\newline 1999.
\newblock Filtering via simulation: {{Auxiliary}} particle filters.
\newblock {\em Journal of the American statistical association},
  94(446):590--599.

\bibitem[\protect\astroncite{Polson and Stroud}{2003}]{polson2003}
Polson, N. and J.~Stroud\leavevmode\nopagebreak\newline 2003.
\newblock Bayesian {{Inference}} for {{Derivative Prices}}.
\newblock In {\em Bayesian {{Statistics}} 7}, Pp.~ 641--650.
\newblock {Oxford University Press}.

\bibitem[\protect\astroncite{Storvik}{2002}]{storvik2002particle}
Storvik, G.\leavevmode\nopagebreak\newline 2002.
\newblock Particle filters for state-space models with the presence of unknown
  static parameters.
\newblock {\em IEEE Transactions on signal Processing}, 50(2):281--289.

\bibitem[\protect\astroncite{Warty et~al.}{2016}]{warty2016}
Warty, S., H.~Lopes, and N.~Polson\leavevmode\nopagebreak\newline 2016.
\newblock Sequential {{Bayesian}} learning for stochastic volatility with
  variance-gamma jumps in returns (with discussion).
\newblock {\em Applied Stochastic Models in Business and Industry}, to appear.

\bibitem[\protect\astroncite{Yun}{2014}]{yun2014out}
Yun, J.\leavevmode\nopagebreak\newline 2014.
\newblock Out-of-sample density forecasts with affine jump diffusion models.
\newblock {\em Journal of Banking \& Finance}, 47:74--87.

\end{thebibliography}

\section*{Appendix A: Particle Filtering Methods}

State space models (\cite{carlin1992monte,duffie1996state,johannes2010}) are central
to inference in financial econometrics. Particle filtering methods are designed to provide state 
inference (\cite{gordon1993novel,carpenter1999improved,pitt1999filtering,storvik2002particle,carvalho2010particle}).

Let $y_t$ denote the data, and $\theta_t$ the state variable. For example, in Merton's jump stochastic 
volatility model, the state variable is $ (J_t , Z_t, V_t)$, corresponding to  the jump times and sizes 
and stochastic volatility. Let $\phi$ denote the  unknown parameters relating to the dynamics of the 
underlying jump and stochastic volatility distributions. For the moment, we suppress the conditioning 
on the parameters $\phi$. We now show how a PF algorithm updates state variables. 

We can factorize  the joint posterior distribution of the data and state variables both ways as
\begin{align*}
p(y_{t+1},\theta_{t+1}|\theta_t) &=p(y_{t+1}|\theta_{t+1})p(\theta_{t+1}|\theta_t)\\
&=p(y_{t+1}|\theta_t)\,p(\theta_{t+1}|\theta_t,y_{t+1}) 
\end{align*}

The goal is to obtain the new filtering distribution $p(\theta_{t+1}|y^{t+1})$ from the current 
$p(\theta_{t}|y^{t})$. A particle representation of the previous filtering distribution is a 
random histogram of draws. We denote this by
$$
p^N(\theta_t|y^{t})=\frac{1}{\,N\,}\sum_{i=1}^N\delta_{\theta_t^{(i)}}.
$$ 
where $ \delta $ is a Dirac measure. As the number of particles increases $N\rightarrow \infty$ the law of large numbers guarantees that this distribution converges to the true filtered distribution $p(\theta_{t}|y^{t})$.

In order to provide random draws of the next distribution, we first resample $\theta_t$'s using the smoothing distribution obtained by  Bayes rule.
$$
p(\theta_t|y^{t+1})\propto p(y_{t+1}|\theta_t)p(\theta_t|y^{t})
$$ 
 Thus, we draw $\theta^{k(i)}_t$ via an
index $k(i)$ from a multinomial with weights
$$
w^{(i)}_t=\frac{p(y_{t+1}|\theta_t^{(i)})}{\sum_{j=1}^Np(y_{t+1}|%
	\theta_t^{(j)})}.$$
We set $\theta^{(i)}_t=\theta^{k(i)}_t$ and {\em ``propagate''} to the next time $t+1$ using
$$p(\theta_{t+1}|y_{1:t+1})=\int
p(\theta_{t+1}|\theta_t,y_{t+1})p(\theta_t|y_{1:t+1})d\theta_t.
$$
Given a particle approximation $ \{ \theta^{(i)} : 1 \leq i \leq N \} $ to
$p^{N}\left(  \theta_t|y^{t}\right)  $, we can use Bayes rule to write
\begin{align*}
p^{N}\left(  \theta_{t+1}|y^{t+1}\right)   &  \propto\sum_{i=1}^{N}p\left(
y_{t+1}|\theta_t^{\left(  i\right)  }\right)  p\left(  \theta_{t+1}|\theta_t^{\left(
	i\right)  },y_{t+1}\right) \label{Mixture2}\\
&  =\sum_{i=1}^{N}w_{t}^{\left(  i\right)  }p\left(  \theta_{t+1}|\theta_t^{\left(
	i\right)  },y_{t+1}\right)  \text{,}%
\end{align*}
where the particle weights are given by
\[
w_{t}^{\left(  i\right)  }=\frac{p\left(  y_{t+1}|\theta_t^{\left(  i\right)
	}\right)  }{\sum_{i=1}^{N}p\left(  y_{t+1}|\theta_t^{\left(  i\right)  }\right)
}\text{.}%
\]
This mixture distribution representation leads to a simple simulation approach for propagating particles to the next filtering distribution. 

The algorithm consists of two steps:
\begin{align*}
&  \text{Step 1. (Resample) Draw }\theta_t^{\left(  i\right)  }\sim
Mult_{N}\left(  w_{t}^{\left(  1\right)  },...,w_{t}^{\left(  N\right)
}\right)  \text{ for }i=1,...,N\\
\text{ }  &  \text{Step 2. (Propagate) Draw }\theta_{t+1}^{\left(  i\right)  }\sim
p\left(  \theta_{t+1}|\theta_t^{\left(  i\right)  },y_{t+1}\right)  \text{ for
}i=1,...,N.
\end{align*}

To implement this algorithm, we need the predictive likelihood for the next observation, $y_{t+1}$, 
given the current state variable $ \theta_t $. It is  defined by
\[
p\left(  y_{t+1}|\theta_t\right)  =\int p\left(  y_{t+1}|\theta_{t+1}\right)  p\left(
\theta_{t+1}|\theta_t\right)  d\theta_{t+1}.%
\]
We also need the  conditional posterior for the next states $ \theta_{t+1} $ given $ (\theta_t , y_{t+1} )$,  given by
\[
p\left(  \theta_{t+1}|\theta_t,y_{t+1}\right)  \propto p\left(  y_{t+1}|\theta_{t+1}%
\right)  p\left(  \theta_{t+1}|\theta_t\right) \; .
\]

Our algorithm has several practical advantages. First,  it does not suffer from the problem of particle 
degeneracy which plagues the standard sample-importance resample filtering algorithms. This effect is 
heightened when $y_{t+1}$ is an outlier. Second, it can easily be extended to incorporate sequential 
parameter learning. It is common to also require learning about other unknown static parameters, denoted by 
$\phi$. To do this, we assume that there exists a conditional sufficient statistic $s_t$ for $\phi$
at time$\,t$, namely
$$
p( \phi | \theta_{1:t} , y_{1:t} ) = p( \phi | s_t )
$$
where $ s_t = s(  \theta_{1:t} , y_{1:t} ) $.
Moreover, we can propagate these sufficient statistics by the deterministic recursion $s_{t+1}=S(s_t,\theta_{t+1},y_{t+1})$
In the next section we develop the sufficient statistics and appropriate recursions for Merton's jump stochastic volatility model.
This will lead to efficient inference for all model parameters.

Given particles $(\theta_t,\phi,s_t)^{(i)},$ $i=1,\ldots,N$. First,  we
resample $(\theta_t,\phi,s_t)^{k(i)}$ with weights
proportional to $p(y_{t+1}| ( \theta_t,\phi)^{k(i)})$. Then we propagate to the next 
filtering distribution
$p(\theta_{t+1}|y_{1:t+1})$ by drawing $\theta^{(i)}_{t+1}\,$from
$p(\theta_{t+1}|\theta^{k(i)}_t,\phi^{k(i)},y_{t+1}),\,i=1,\ldots,N$. We next
update the sufficient statistic for $i=1,\ldots,N$,
$$
s_{t+1}=S(s_t^{k(i)},\theta^{(i)}_{t+1},y_{t+1}),
$$ 
This represents a deterministic propagation. Parameter learning is
completed by drawing $\phi^{(i)}$ using $p(\phi|s^{(i)}_{t+1})$ for
$i=1,\ldots,N$. We now gives specifics of the algorithm for two models, the Merton model
with constant and stochastic volatility.
We now track the state, $\theta_t$, and conditional sufficient statistics, $s_t$, which will be used to perform off-line learning for $\phi$.

The algorithm now consists of four steps:
\begin{align*}
&  \text{Step 1. (Resample) Draw Index } k_t(i) \sim
Mult_{N}\left(  w_{t}^{\left(  1\right)  },...,w_{t}^{\left(  N\right)
}\right)  \text{ for }i=1,...,N\\
& \text{The weights are proportional to } p( y_{t+1} | ( \theta_t , s_t )^{(i)} ) \\
\text{ }  &  \text{Step 2. (Propagate) Draw }\theta_{t+1}^{\left(  i\right)  }\sim
p\left(  \theta_{t+1}| (\theta_t , s_t)^{k \left(  i\right)  },y_{t+1}\right)  \text{ for
}i=1,...,N.\\
\text{ }  &  \text{Step 3. (Update) Deterministic } s_{t+1}^{\left(  i\right)  }=
S\left( s_t^{k(i)} , \theta_{t+1}^{(i)}  ,y_{t+1}\right)  \text{ for
}i=1,...,N.\\
\text{ }  &  \text{Step 4. (Learning) Offline } \phi^{\left(  i\right)  }\sim
p\left(  \phi | s_{t+1}^{(i)} \right)  \text{ for
}i=1,...,N.\end{align*}
\end{document}